\pdfoutput=1

\documentclass[aps,prb,amsmath,amssymb,floatfix,twocolumn,amsmath,superscriptaddress,twocolumn,nofootinbib,tighten,letterpaper]{revtex4-2}

\usepackage[utf8]{inputenc}
\usepackage{bm}
\usepackage{siunitx}

\usepackage{enumerate}
\usepackage{amsfonts}
\usepackage{amsmath}
\usepackage{amssymb}
\usepackage{color}
\usepackage{soul}
\usepackage{float}
\usepackage{bm}
\usepackage{graphicx}
\usepackage{graphics}

\usepackage{graphicx}
\usepackage[utf8]{inputenc}
\usepackage{physics}
\usepackage{verbatim}
\usepackage[colorlinks,allcolors=blue]{hyperref}
\usepackage{tikz-feynman}
\usepackage{xcolor,soul}
\usepackage{cancel}

\usepackage{tikz}
\usetikzlibrary{angles, quotes, intersections}

\newcommand{\bs}[1]{\boldsymbol{#1}}

\usepackage[colorlinks,allcolors=blue]{hyperref}

\let\propto=\sim
\let\epsilon=\varepsilon

\renewcommand{\vec}[1]{\bm{#1}}

\definecolor{DarkRed}{rgb}{0.80,0,0}
\definecolor{DarkGray}{rgb}{0.7,0.7,0.7}

\newcommand{\prlsection}[1]{\textit{#1}.\kern0.05em---\kern0.05em\ignorespaces}
\usepackage[capitalise]{cleveref}

\begin{document}

\title{Axionic quantum criticality of generalized Weyl semimetals}

\author{Gabriel Malav\'e}
\affiliation{Facultad de Física, Pontificia Universidad Católica de Chile, Vicuña Mackenna 4860, Santiago 8331150, Chile}

\author{Rodrigo Soto-Garrido}
\affiliation{Facultad de Física, Pontificia Universidad Católica de Chile, Vicuña Mackenna 4860, Santiago 8331150, Chile}

\author{Vladimir Juri\v{c}i\'c}
\affiliation{Departamento de F\'isica, Universidad T\'ecnica Federico Santa Mar\'ia, Casilla 110, Valpara\'iso, Chile}

\author{Bitan Roy}
\affiliation{Department of Physics, Lehigh University, Bethlehem, Pennsylvania, 18015, USA}

\begin{abstract}
We formulate a field-theoretic description for $d$-dimensional interacting nodal semimetals, featuring dispersion that scales with the linear and $n$th power of momentum along $d_L$ and $d_M$ mutually orthogonal directions around a few isolated points in the reciprocal space, respectively, with $d_L+d_M=d$, and residing at the brink of isotropic insulation, described by $N_b$-component bosonic order parameter fields. The resulting renormalization group (RG) procedure, tailored to capture the associated quantum critical phenomena, is controlled by a ``small" parameter $\epsilon=2-d_M$ and $1/N_f$, where $N_f$ is the number of identical fermion copies (flavor number) when in conjunction $d_L=1$. When applied to three-dimensional interacting general Weyl semimetals ($d_L=1$ and $d_M=2$), characterized by the Abelian monopole charge $n>1$, living at the shore of the axionic insulation ($N_b=2$), a leading-order RG analysis suggests the Gaussian nature of the underlying quantum phase transition, around which the critical exponents assume mean-field values. A traditional field-theoretic RG analysis yields the same outcomes for simple Weyl semimetals ($n=1$, $d_L=3$, and $d_M=0$). Consequently, emergent marginal Fermi liquids showcase only logarithmic corrections to physical observables at intermediate scales of measurements.
\end{abstract}
\maketitle


\emph{Introduction}.~Fermi liquids, prominent itinerant electronic systems, host gapless sharp fermionic excitations around closed contours in the momentum space, known as Fermi surfaces~\cite{landau1957theory}. In the presence of electronic interactions, they become unstable toward charge density wave, ferromagnet, antiferromagnet, and nematicity~\cite{Sachdevbook, MetlitskiSachdev2010a, MetlitskiSachdev2010b, SSLee2013, Sur-Lee2016, BergPRX2016, BergPRX2017}, for example, ultimately giving way to superconductivity at the lowest temperature~\cite{KohnLuttinger1965, Baranov1992, ShankarRMP1994}. These orderings set in through quantum phase transitions (QPTs), yet their controlled field-theoretic description remains challenging due to an infinite number of fermionic nodes in the system, coupled to critical bosonic fluctuations.

Nodal Fermi liquids also feature gapless fermionic excitations, but around a few special points in the Brillouin zone. Besides showcasing novel topological responses~\cite{ArmitageRMP2018}, they constitute a key testbed for itinerant quantum critical properties. Dirac and Weyl liquids, with the chemical potential pinned at the band touching points at half filling, are their prime candidates. The linear energy-momentum relation, endowing the elegant Lorentz invariance, allows traditional field-theoretic renormalization group (RG) techniques to capture the underlying quantum criticality therein~\cite{ZinnJustinbook, Rosenstein1993, GRACEY1994, VojtaSachdev2000, Moshe2003, SSLee2007PRB, HJR2009, HJV2009, BR2011, Maciejko2016PRB, RoyKennett2018}. In parallel, field-theoretic predictions can be tested from infamous sign-problem-free quantum Monte Carlo simulations~\cite{Sorella1992, HerbutAssad2013, Toldinetal2015, Sorella2016, Li2015, Scalettar2018, LangLauchli2019, Scalettar2020, Assaad2023}. Fascinatingly, these methodologies apply to some open or non-Hermitian Dirac systems~\cite{JuricicRoyNH2024, MurshedNH2024, YuNH2024}.

However, established traditional RG techniques that tacitly assume a space-time Lorentz symmetry~\cite{ZinnJustinbook, peskin1995} are inadequate when topological semimetals display \emph{nonlinear} dispersion along certain high-symmetry directions due to the lack of Lorentz symmetry to capture quantum criticality therein~\cite{sur2019}. This is the case in Weyl semimetals (WSMs) with the underlying Weyl nodes characterized by Abelian monopole charge $n>1$, a family of systems at the central focus in this article, hereafter named general WSMs. The fermionic dispersion therein is linear along the direction connecting two Weyl nodes ($k_z$), while in the perpendicular plane, it scales as $k^n_\perp$, where $\vec{k}_\perp=(k_x,k_y)$~\cite{XuPRL2011, FangPRL2012, Yang2014, Huang2016, Liu2017}. Vanishing density of states $\varrho(E) \sim |E|^{2/n}$ around the Weyl nodes protects them against sufficiently weak electron-electron interactions. Nevertheless, an axionic insulator emerges as the dominant ground state for strong repulsive electronic interactions, in which two Weyl nodes get coupled and the translational symmetry is broken spontaneously~\cite{WangZhang2013, MaciejkoNandkishore2014, WangPeng2016, Laubach2016, roy2017}.

The associated U(1) Goldstone-like sliding mode (with a \emph{small} mass) is analogous to the illusive pseudoscalar light-mass bosonic axion particle, proposed four decades ago in the context of high-energy physics~\cite{PecciQuinn1977, Weinberg1978, Wilczek1978}. Besides axion electrodynamics~\cite{Wilczek1987}, a line vortexlike defect in the axion insulator supports $n$ unidirectional current carrying modes through its core, when a magnetic field is applied in the direction of the Weyl point separation, manifesting the Callan-Harvey mechanism~\cite{Callan1985, RoySau2015}.

Here, we formulate a general field-theoretic RG approach to capture the quantum critical behavior of a $d$-dimensional semimetal, in which the fermionic dispersion scales with the linear  and $n$th power of momentum along $d_L$ and $d_M$ mutually orthogonal directions, respectively, with $d_L+d_M=d$, near a QPT to an insulator, described by an $N_b$-component bosonic order-parameter field. Naturally, this procedure encompasses a  three-dimensional (3D) general WSM ($d_L=1$ and $d_M=2$) to an axionic insulator ($N_b=2$) QPT. The proposed RG approach is controlled by a `small' parameter $\epsilon=2-d_M$ and $1/N_f$, where $N_f$ is the number of identical fermion copies (flavor number) when $d_L=1$. From a leading-order RG analysis we find that such a QPT in any general WSM is Gaussian in nature. For conventional WSM ($n=1$ and $d_L=3$), we arrive at this conclusion from the existing RG procedure. Thus, the critical exponents, which ultimately assume their mean-field values, only \emph{logarithmically} influence the scaling of various physical observables at intermediate scales (temperature or frequency) of measurements in the emergent critical marginal Fermi liquids.

This framework should be contrasted with the existing literature focusing on two-dimensional anisotropic semimetals (2D-ASMs) involving, at most, quadratic dispersion ($n=2$) with $d_L=d_M=1$ in the presence of long-range Coulomb interaction~\cite{NagaosaPRL2016}. Our approach should also be compared with RG studies on such 2D-ASMs near dynamic mass generation via spontaneous symmetry breaking, by employing either an $\epsilon$ expansion~\cite{sur2019}, similar to the spirit of the present work, or a $1/N_f$ expansion in terms of a large number of fermion flavors ($N_f$)~\cite{UchoaPRB2019, ChristouPRR2020}. Similar dynamical mass generation has also been addressed in three-dimensional general Weyl semimetals (with an arbitrary $n$ and $d_L=d_M/2=1$) and 2D-ASMs by considering only local four-fermion interactions, thus ignoring effects of order parameter fluctuations, by treating $1/n$ (determining the band curvature) as the expansion parameter~\cite{roy2017, RoyFosterPRX2018}. We note that such a $1/n$ expansion has also been employed to study the effects of quench disorder near the Weyl semimetal to band insulator quantum critical point (QCP) in $d=3$ where $d_L/2=d_M=1$~\cite{RoySlagerJuricicPRX, Ohtsuki2018PRB}. While the present work unfolds distinct emergent quantum criticality close to the spontaneous mass generation in point-node semimetals, featuring dispersion that scales with the $n$th power of momentum along the non-relativistic directions, close to the spontaneous mass generation, the associated theoretical approach shares some unavoidable overlap with the existing ones in the literature~\cite{NagaosaPRL2016, sur2019, UchoaPRB2019, ChristouPRR2020}.

\emph{Model}.~The effective single-particle Hamiltonian describing such a non-interacting nodal-point semimetal is 
\allowdisplaybreaks[4] 
\begin{equation}~\label{eq:generalHamil}
    H_0\left(k_i\right)=\sum_{l=1}^{d_L} v_{_l} \Gamma_l k_l 
		+ \sum_{n=1}^{d_M} \Gamma_{d_L+n} \varepsilon_n \left(k_{d_L+1}, \ldots, k_d\right),
\end{equation}
where $k_i$ are the components of momentum, $\{\Gamma_i\}$ is a set of mutually anticommuting Hermitian matrices, satisfying $\{\Gamma_i,\Gamma_j\}=2 \delta_{ij}$, $\varepsilon_n$ are nonlinear functions of their arguments, $\delta_{ij}$ is the Kronecker delta function, and $v_{_l}$ bears the dimension of Fermi velocity. For simplicity, we set $v_{_l}=v=1$ for all $l$. Short-range Hubbard-like interactions can trigger a symmetry breaking by generating a (multicomponent) mass term $\sum_{j=1}^{N_b} M_j \Upsilon_j$, where $M_j =\langle \Psi^\dagger \Upsilon_j \Psi \rangle$ and $\{\Upsilon_j\}$ is another set of mutually anticommuting Hermitian matrices such that $\{\Upsilon_i,\Upsilon_j\}=2 \delta_{ij}$ and $\{\Gamma_i,\Upsilon_j\}=0$. The internal structure of fermionic spinors ($\Psi^\dagger$ and $\Psi$) depends on the microscopic details, which we do not delve into here. Such an ordering leads to a fully and isotropic gapped state (insulator), where the magnitude of $\vec{M}$ determines the condensation energy gain. The dimensionality of all $\Gamma_j$s and $\Upsilon_j$s is $4N$.

Near the QCP, relevant degrees of freedom are gapless nodal fermions and bosonic order-parameter fluctuations, coupled via a Yukawa interaction. The effective imaginary-time action reads
\begin{widetext}
\begin{align}~\label{eq:action}
    S =  \sum^{N_f}_{p=1} \left[\int_k \psi_{p, k}^{\dagger} G_0^{-1}(k) \psi_{p, k}
		+ \frac{g_{_0}}{\sqrt{N_f}} \int_{k, q} \bs{\phi}_q \cdot  \psi_{p, k+q}^{\dagger} \bs{\Upsilon} \psi_{p, k} \right]
		+\frac{1}{2}  \int_q D_0^{-1}(q) \bs{\phi}_{\bar{q}} \cdot\bs{\phi}_q 
    +\frac{u_{_0}}{2}\int_{k, q, q'} \bs{\phi}_{\bar{q}} \cdot\bs{\phi}_{q+q'}\,\,\bs{\phi}_{\bar{k}} \cdot\bs{\phi}_{k-q'},
\end{align}
\end{widetext}
where $\bar{x}=-x$, $k=(k_0,k_1,\cdots,k_d)$, with $k_0$ being the Euclidean frequency, while $ \psi_{p, k}$ and $\bs{\phi}_q$ represent the $p$th copy of the Grassmann and order parameter fields, respectively. The bare Yukawa and $\phi^4$ couplings are $g_{_0}$ and $u_{_0}$, respectively. Here, we consider $N_f$ identical copies of $4N$-component fermions. The bare fermionic and bosonic propagators are given by 
\begin{eqnarray}
&&G_0(k)=\left[i k_0 + H_0(k_1,...,k_d)\right]^{-1},\\
&&\text{and} \: D_0(q)=\left[c^2\left(q_0^2+\sum_{l=1}^{d_L} q_l^2\right)+\sum_{n=d_L+1}^{d_M} q_n^2+|\vec{M}|^2\right]^{-1}, \nonumber \\
\end{eqnarray}
respectively, with the parameter $c$ encoding the anisotropy in the bosonic dynamics, stemming from the anisotropic dispersion of the fermionic excitations.

\emph{Scaling}.~For the scaling analysis, we conveniently define a $(d_L+1)$-dimensional frequency-momentum vector $\bs{k}\equiv(k_0,k_1,\cdots,k_{d_L})$ and $d_M$-dimensional momentum $\bs{K}\equiv(k_{d_L+1},\cdots,k_{d})$, with the scaling dimensions $[\bs{K}]=1$ and $[\bs{k}]=z_{\bs{k}}$. When the function $\varepsilon_n$ from Eq.~\eqref{eq:generalHamil} depends on the $n$th power of its arguments in the $d_M$-dimensional hyperplane, as is the case for general WSMs (discussed shortly), the action in Eq.~\eqref{eq:action} is invariant under the tree-level scaling upon setting $z_{\bs{k}}=n$ at the non-interacting fixed point. The fermionic action yields $[\psi] = -[z_k(d_L+2)+d_M]/2$ as $[G_0^{-1}]=z_{\bf k}$. Furthermore, we have $[c] = 1-z_{\bs{k}}$ and $[M] = 1$, leading to the scaling dimension of the bosonic field $[\phi^2] = -2-[z_k(d_L+1)+d_M]$, since $[D_0^{-1}] = 2$. Then for the Yukawa term we find $[g_{_0}^2] = 2-[z_{\bf k}(d_L-1)+d_M]$, which is marginal on the line $z_{\bs{k}}(d_L-1)+d_M=2$, and passes through the point $(d_L,d_M)=(1,2)$, representing the marginality condition in the non-Lorentz symmetric theory.

Interacting fixed points where the Yukawa coupling is dimensionful can then be reached by independently tuning the deviations from this marginal line in term of the `small' parameters $\epsilon_L=\bar{d}_L-d_L$ and $\epsilon_M=\bar{d}_M-d_M$, where $z_{\bs{k}}(\bar{d}_L-1)+\bar{d}_M=2$. This is accomplished through dimensional regularization within the minimal subtraction scheme~\cite{ZinnJustinbook, peskin1995}, yielding the RG flow equation for the Yukawa coupling. On this line the quartic bosonic coupling is irrelevant for any $n>1$ as $[u_{_0}] = 4 - z_{\bf k}(d_L+1) - d_M$, which we do not therefore consider in the following analysis. By contrast, a Lorentz symmetric theory is marginal at the point $(d_L,d_M)=(3,0)$, where both $g_{_0}$ and $u_{_0}$ are marginal and the access to the nontrivial QCP is controlled by an $\epsilon=3-d$ expansion.  

\begin{figure}[t!]
\includegraphics[width=1.00\linewidth]{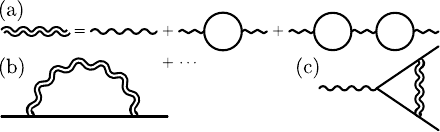}
\caption{(a) Feynman diagram representation of the dressed boson propagator (double wavy line) resulting from the random phase approximation processes, where the solid (single wavy) line correspond to fermion (bare boson) propagator. Feynman diagrams for (b) fermionic self-energy and (c) Yukawa vertex corrections through dressed boson propagator.              
}~\label{fig:Fig1}
\end{figure}

\emph{Interacting WSM}.~With the stage being set, we proceed to capture the emergent quantum criticality in general WSMs,
with their Hamiltonian given by 
\begin{align}~\label{eq:H}
 H_{n}(k_\perp,k_z)= \alpha_n k_{\perp}^n\left[\Gamma_1 C_{n \phi_{\bs{k}}} + \Gamma_2 S_{n \phi_{\bs{k}}} \right] +\Gamma_3 v_z k_z, 
 \end{align}
when they reside at the brink of axionic insulation, where $C_x=\cos(x)$ and $S_x=\sin(x)$. Mutually anticommuting  $\Gamma$ matrices are $\Gamma_1= \tau_0 \otimes \sigma_1$, $\Gamma_2= \tau_0 \otimes \sigma_2$, and $\Gamma_3=\tau_3 \otimes \sigma_3$, where $\otimes$ stands for Kronecker product. Pauli matrices $\sigma_\mu$ ($\tau_\mu$) act on the orbital (valley) subspace, and $\phi_{\bs{k}}=\tan^{-1}(k_y/k_x)$. The parameters $\alpha_n$ and $v_z$ are real and positive, with $\alpha_1$ and $\alpha_2$ bearing the dimension of Fermi velocity and inverse mass, respectively. The bare fermionic propagator reads $G_0(\bs{k},\bs{K})=[ i k_0 + H_{n}(k_\perp,k_z)]^{-1}$, where the  corresponding momentum 2-vectors are $\boldsymbol{k} \equiv (k_0,k_z)$ and $\boldsymbol{K} \equiv (k_x,k_y)$.

For sufficiently strong interactions, WSMs with any $n$ can undergo a QPT into a translational symmetry breaking axion insulator~\cite{roy2017}, which is two-component ($XY$) order with $N_b=2$ and is represented by the mutually anticommuting matrices $\Upsilon_1 \equiv \Gamma_4 = \tau_1 \otimes \sigma_3$ and $\Upsilon_2 \equiv \Gamma_5 = \tau_2 \otimes \sigma_3$. These two matrices also anticommute with the $\Gamma$ matrices in the Hamiltonian of the WSM [Eq.~\eqref{eq:H}]. Therefore, axionic insulator fully gaps out the nodal fermionic excitations. The corresponding effective single-particle Hamiltonian is 
\begin{equation}
H_{\rm AI}= |\vec{M}| \; \left[ \Upsilon_1 \cos(\theta) + \Upsilon_2 \sin(\theta) \right].
\end{equation}
The U(1) Goldstone mode describing the fluctuations of $\theta$ is known as the axion field, and it couples with the external electromagnetic fields through the axion term $\theta {\bf E} \cdot {\bf B}$. In quantum crystals, $\theta$ describes the sliding mode.

\emph{RG analysis}.~The $4N$-dimensional $\Gamma$ and $\Upsilon$ matrices allow us to perform the RG analysis near the QCP associated to a general $N_b$-component insulating order parameter, with our ultimate focus near the axionic QCP. Before delving into such an exercise in general WSMs, we consider simple WSM with Lorentz invariance, $d_L=3$, and $d_M=0$ in Eq.~\eqref{eq:action}. The coupled RG flow equations can then be readily obtained~\cite{RJH-PRB2013} and are given by
\begin{align}
\frac{dg^2}{d\ell}&=\epsilon g^2 -\left(N+\frac{1}{N_f}\right)\frac{g^4}{8\pi^2},\\
\frac{du}{d\ell}&=\epsilon u -\left(u-\frac{Ng^2}{N_f}\right)\frac{g^2}{4\pi^2}-\left(4+\frac{N_b}{2}\right)\frac{u^2}{8\pi^2},
\end{align}
which feature a nontrivial QCP for $d<3$ ($\epsilon>0$) at $g_\star\sim \epsilon$ and $u_\star \sim \epsilon$. Here, $\ell \equiv -\ln \mu$ is the logarithm of the running RG scale and $\mu$ is the momentum scale. Such a QCP is characterized by the non-mean-field correlation length exponent, $\nu^{-1}=2-(N_b+2) u_\star/(16\pi^2)-N g_\star^2/(8\pi^2)$,  and the anomalous dimensions of the fermionic and  bosonic fields,  $\eta_\psi=g_\star^2/(16\pi^2N_f)$  and $\eta_\phi=N g_\star^2/(8\pi^2)$, respectively, to the leading order in the $\epsilon$ expansion.

Next, we focus on 3D general WSMs and cast the inverse fermionic propagator as $G_0^{-1}(\bs{k},\bs{K})= i\Gamma_{13} \left[\bs{k}\cdot\bs{\gamma}+|\bs{K}|^n \hat{\bs{n}}_{\bs{k}} \cdot \bs{\Gamma} \right]$, where $\bs{\gamma}=(\Gamma_{13},-\Gamma_{20})$, $\bs{\Gamma}=(\Gamma_{12},-\Gamma_{11})$, $\Gamma_{ij}=\tau_i \otimes \sigma_j$, $\hat{\bs{n}}_{\bs{k}} = (\cos(n\phi),\ \sin(n\phi))$, and $\phi\equiv\phi_{\bs{k}}$. We perform the RG analysis in two steps. First, we calculate the bosonic self-energy diagram within the $\epsilon_M=2-d_M \equiv \epsilon$ expansion around the marginal condition ${d}_M=2$, keeping $d_L=1$ fixed. This procedure explicitly takes into account the anisotropy of the fermionic dispersion~\cite{sur2019}. The resulting dressed bosonic propagator is obtained by resumming the dominant random phase approximation processes [Fig.~\ref{fig:Fig1}(a)], eliminating the putative infrared (IR) divergences in the limit $c \to 0$, which takes the form (see Supplemental Material~\cite{SM})
\begin{equation}
 \label{eq:dressed-bosonic-prop}
 D^{-1}(\bs{q},\bs{Q})=N\left(|\bs{Q}|^2+ g^2 A_n|\bs{q}|^{2/n}\right),
 \end{equation}
 where $g \equiv g_{_0} \mu^{(d_M-2)/2}$ is the dimensionless Yukawa coupling. Upon setting the bosonic mass to zero, we find
 \begin{equation}
 \label{eq:An}
    A_n = -\frac{1}{(2 \pi)^2} \int_0^{\infty} dz \int_0^1 dx \frac{x^2-z^n}{x(x-1)-z^n}.
\end{equation}
At an intermediate momentum scale $\mu$, the action from Eq.~\eqref{eq:action} then takes the form (see Supplemental Material~\cite{SM})
\begin{widetext}
\allowdisplaybreaks[4]
\begin{align}\label{eq:action_mu}
    S= \sum_{p=1}^{N_f} \left[ i \int_k \psi^{\dagger}_{p, k} \Gamma_{13} \left[\bs{k}\cdot\bs{\gamma}+|\bs{K}|^n\, \hat{\bs{n}}_{\bs{k}}  \cdot \bs{\Gamma} \right] \psi_{p, k}
		+\frac{g \mu^{\frac{2-d_M}{2}}}{\sqrt{N_f}} \int_{k, q} \psi^\dagger_{p, k+q} \widetilde{\Phi}_q \psi_{p, k} \right]
    +\frac{1}{8} \int_q\left(|\boldsymbol{Q}|^2+M^2\right) \operatorname{tr}\left\{\widetilde{\Phi}_{\bar{q}} \widetilde{\Phi}_q\right\},
\end{align}
where we keep only the marginal and relevant terms close to $d_L=1$ and $d_M=2$, and $\widetilde{\Phi}_q = \sum^{N_b}_{j=1}\phi_j(q) \Upsilon_j$.

This procedure brings the action into a renormalizable form which, in the second step, we treat by adding counterterms. After setting the renormalized bosonic mass $M$ to zero (critical plane), the renormalized action reads
\allowdisplaybreaks[4] 
\begin{align}\label{eq:action_ren}
   S_{r}=i \int_k \psi^{\dagger}_{k} \left[
    \mathcal{Z}_1 \bs{k}\cdot\bs{\gamma} + \mathcal{Z}_2 |\bs{K}|^n \hat{\bs{n}}_{\bs{k}} \cdot \bs{\Gamma} \right] \psi_{k}
  +\frac{1}{8} \int_q \mathcal{Z}_3|\boldsymbol{Q}|^2 \operatorname{tr}\left\{\widetilde{\Phi}_{\bar{q}} \widetilde{\Phi}_q\right\}+g \mu^{(3-d) / 2}\mathcal{Z}_4 \int_{k, q} \psi^\dagger_{k+q} \widetilde{\Phi}_q \psi_{k}. 
\end{align}
\end{widetext}
To the leading order in the $\epsilon$ expansion, $\mathcal{Z}_n=1+Z_n/\epsilon$ are renormalization factors, absorbing the ensuing divergences in the $\epsilon=2-d_M$ expansion close to $d_M=2$. We emphasize that here we use the dressed (resummed) damped bosonic propagator, given by Eq.~\eqref{eq:dressed-bosonic-prop}, to regulate the putative IR divergences in the fermionic self-energy [Fig.~\ref{fig:Fig1}(b)] and vertex [Fig.~\ref{fig:Fig1}(c)] corrections in the limit $c \to 0$~\cite{sur2019}. The explicit calculation of the renormalization factors from the fermionic self-energy yields $Z_1=-N_bg^2/(8\pi^2N N_f)$ and $Z_2= (n N_b g^2/(8\pi^2N N_f))\ln(C_ng^2)$, where $C_n$ is a real function of $n$ (see Supplemental Material~\cite{SM}). From the bosonic self-energy, we find $Z_3=-Nn^2 g^2/(4\pi^2)$. The Yukawa vertex correction gives $Z_4=(N_b-2)(ng^2/8\pi^2N N_f)\ln(A_ng^2)$, which vanishes for the axionic $XY$ order parameter ($N_b=2$)~\cite{Sur-Lee2016}.

 The relations between the bare (with the subscript `$B$') and renormalized quantities are given by
\allowdisplaybreaks[4]  
\begin{align}
  &  \bs{k}_B = \mathcal{Z}_{\bs{k}}^{-1} \bs{k}, 
    \bs{K}_B = \bs{K}, \psi_B=\sqrt{\mathcal{Z}_{\psi}}\psi, \phi_B=\sqrt{\mathcal{Z}_{\phi}}\phi,\nonumber\\  
  &  g_{_0} \equiv g_{_B} = \mu^{\frac{2-d_M}{2}} \frac{\mathcal{Z}_{\bs{k}}^4\mathcal{Z}_4}{\mathcal{Z}_{\psi}\sqrt{\mathcal{Z}_{\phi}}}g,
\end{align}
where $\mathcal{Z}_{\bs{k}}=\mathcal{Z}_2/\mathcal{Z}_1$, $\mathcal{Z}_{\psi}=\mathcal{Z}_2\mathcal{Z}_{\bs{k}}^2$ and $\mathcal{Z}_{\phi}=\mathcal{Z}_3\mathcal{Z}_{\bs{k}}^2$. The anisotropy exponent $z_{\bs{k}}$ and the anomalous dimensions ($\eta_\psi$ and $\eta_\phi$) of the fields are defined as
\begin{equation}
    z_{\bs{k}}=n-\frac{\partial \ln \mathcal{Z}_{\bs{k}}}{\partial \ln\mu}, \;  
		\eta_{\psi}=\frac{1}{2}\frac{\partial \ln \mathcal{Z}_{\psi}}{\partial \ln \mu}, \;  
		\eta_{\phi}=\frac{1}{2}\frac{\partial \ln \mathcal{Z}_{\phi}}{\partial \ln \mu}.
\end{equation}
The RG flow equation  for the Yukawa coupling is obtained from the condition of the bare coupling being fixed, $\partial g_{_B}/\partial \ln\mu=0$. After solving this system of equations and using the obtained values of the $Z$ factors (see Supplemental Material~\cite{SM}), we find the anomalous dimensions and the RG flow equation for the Yukawa coupling, which we analyze next.

\emph{Results}.~The RG flow equation for the Yukawa coupling reads as
\begin{equation}~\label{eq:RGflowgeneral}
    \frac{d g^2}{d \ell}=\epsilon g^2-\frac{g^4}{4 \pi^2}\biggl[N n^2 + \frac{N_b+n(N_b-2)\ln(e A_n g^2)}{N N_f} \biggr].
\end{equation}
with $A_n$ given in Eq.~\eqref{eq:An}. Therefore, the noninteracting fixed point at $g_\star^2=0$ is IR stable for any $N_b \leq 2$ when $\epsilon=0$. On the other hand, a IR stable fixed point $g_*^2 \propto \epsilon$ is obtained for any $N_b$ and $d_M<2$. In the limit $\epsilon \ll 1$ and $N_f\gg1$, the $XY$ QCP ($N_b=2$) is located at  
\begin{equation}
g_*^2=  \frac{4\pi^2}{Nn^2} \left(1-\frac{2}{N^2 n^2N_f}\right)\epsilon,
\end{equation}
where we find 
\begin{equation}
z_{\bs{k}}= n + \frac{\epsilon\ln\epsilon}{N^2 nN_f}, \;
\eta_{\psi}= -\frac{3 \epsilon\ln\epsilon}{2N^2nN_f}, \;
\eta_{\phi}= \frac{\epsilon}{2} - \frac{\epsilon\ln\epsilon}{N^2nN_f}
\end{equation}
and the correlation length exponent ($\nu$), obtained from the relevant RG flow of the bosonic mass (the tuning parameter for the QPT), is given by 
\begin{equation}
 \nu^{-1}\equiv2-2\eta_{\phi}= 2 - \epsilon + \frac{2}{N^2nN_f}\epsilon\ln\epsilon.
\end{equation}
Therefore, the axionic QCP in an interacting general WSM is located at $g^2_*=0$. At this fixed point, all the exponents ultimately assume mean-field values, namely $z_{\bs{k}}=n$, $\eta_\psi=0$, $\eta_\phi=0$, and $\nu=1/2$. However, these mean-field values are approached only logarithmically slowly as a function of the running scale ($\mu$), yielding a marginal Fermi liquid at the axionic QCP. Therefore, the marginal Fermi liquid showcases logarithmic corrections to the scaling of various physical observables as a function of temperature ($T$) or frequency ($\omega$), playing the role of the running scale, in contrast to those in noninteracting systems, displaying power-law scaling, discussed next.

For example, the specific heat ($C_v$) and compressibility ($\kappa$) scale as $T^{1+d_M/z_{\bs{k}}}$ and $T^{d_M/z_{\bs{k}}}$, respectively, and the Gr\"uneisen ratio scales as $\Gamma_{G} \sim T^{-(2+d_M/z_{\bs{k}})}$. The band curvature of the fermionic dispersion is modified according to $[k^2_z + k_\perp^{2z_{\bs{k}}}]^{1/2}$. While the spectral density function near the Weyl points scales as ${\mathcal A}(\omega) \sim \omega^{-1+ 2 \eta_\psi}$, the dynamic structure factor for the order-parameter fluctuations goes as ${\mathcal S}(\omega) \sim \omega^{-2+ 2 \eta_\phi}$. The scaling of the zero-temperature longitudinal optical conductivity at finite frequency can be obtained from the gauge invariance of the current-current correlator, yielding $\sigma_{zz} \sim \omega^{-1+(2-\epsilon)/z_{\bs{k}}}$ and $\sigma_{jj} \sim \omega^{1-\epsilon/z_{\bs{k}}}$ for $j=x$ and $y$. The solution of Eq.~\eqref{eq:RGflowgeneral}, yielding the running Yukawa coupling $g(\ell)$ determines the exponents controlling the scaling of physical observables as $x \equiv x(g(\ell))$ with $x=z_{\vec{k}}$, $\eta_\psi$, and $\eta_\phi$. Here, $\ell=\ln(T_0/T)$ or $\ln(\omega_0/\omega)$, and $T_0$ ($\omega_0$) is the ultraviolet temperature (frequency) scale. Only near axionic ordering ($N_b=2$), are $g(\ell)$ and observables not affected by the $\ln(g^2)$ term in the RG flow equation.

Before closing this section, we briefly comment on the structure of the fixed point for $N_b\neq 2$. Within the realm of the $\epsilon$ expansion, there exists an IR stable fixed point at $g_*^2= a/W_s(a A_n e^{b})$, where $a=4\pi^2N N_f \epsilon/[n(N_b-2)]$, $b=(N^2 n^2 N_f+N_b+n(N_b-2))/[n(N_b-2)]$. Here $W_s(x)$ is the product of the logarithm on the $s$th branch, and for $N_b=1$ ($N_b \geq 3$), the parameter $s=-1$ ($s=0$) (see Supplemental Material~\cite{SM}). As $g^2_* \sim \epsilon$, this fixed point also manifests Gaussian quantum criticality. However, even if we set $\epsilon=0$ in Eq.~\eqref{eq:RGflowgeneral}, there exists an IR stable fixed point for $N_b \geq 3$ at 
\begin{equation}
g^2_* = \frac{1}{A_n e} \; \exp \left[- \frac{NN_f}{n(N_b-2)} \left( N n^2 + \frac{N_b}{NN_f} \right) \right]
\end{equation}
which is beyond the territory of the $\epsilon$ expansion. For $N_b \neq 2$, the logarithmic dependence of the RG flow equation on the Yukawa coupling leaves its signature on $g(\ell)$ which can be obtained by solving Eq.~\eqref{eq:RGflowgeneral} that, in turn, also affects the scaling of physical observables at an intermediate scale ($\ell$), as their governing exponents ($z_{\vec{k}}$, $\eta_\psi$, and $\eta_\phi$) depend on $g(\ell)$.

\emph{Summary and discussions}.~To summarize, here we expose the quantum critical properties of interacting 3D WSMs (both simple and general) when they reside at the brink of axionic insulation, triggered by suitable short-range Hubbard-like interactions. While in such a situation the traditional field-theoretic RG approach can be applied in simple WSMs, our proposed field-theoretic approach is tailored to address similar quests in generic non-Lorentz symmetric interacting semimetals that encompass general WSMs. Both methods, nonetheless, predict that the WSM to axion insulator QPT is Gaussian in nature, and the critical exponents ultimately acquire their mean-field values. The resulting marginal Fermi liquid, residing around the underlying QCP, thus features logarithmic corrections to the scaling of thermodynamic and transport quantities as a function of temperature and frequency, respectively. Strong Hubbard-like interactions can make WSMs susceptible to the nucleation of rotational symmetry-breaking nematic orders, which, in simple and general WSMs, respectively, shift the location of and split the Weyl nodes~\cite{roy2017, JianYao2017}. Future investigations will unfold critical properties around such orderings.

Unfortunately, signatures of the axionic insulation in Weyl materials have remained elusive so far, unless they are placed in strong magnetic fields~\cite{Gooth2019}. Application of external magnetic fields quenches the kinetic energy and gives birth to Landau levels that are dispersive in the field direction. The resulting finite density of states triggers the onset of axion insulator even for infinitesimal interactions, following the magnetic catalysis mechanism~\cite{RoySau2015, LiRoyDasSarma2016}. However, WSMs realized on strongly correlated quantum crystals, such as 227 pyrochlore iridates with multipolar magnetic orders~\cite{Savrasov2011, YamajiImada2014, WitczakKrempa2014, GoswamiRoyDasSarma2017, LadovrechisMengRoy2021} and Kondo materials~\cite{Lai2017, Chen2022, Dzsaber2017}, can be the ideal place to harness such exotic insulating ground state and hallmarks of emergent marginal Fermi liquids in terms of the logarithmic scaling corrections of physical observables, predicted in this work.

\acknowledgments

~This work was supported by ANID/ACT210100 (G.M., R.S.-G., and V.J.), the Swedish Research Council Grant No.\ VR 2019-04735 (V.J.), and Fondecyt (Chile) Grants No.\ 1241033 (R.S.-G.) and No.\ 1230933 (V.J.). B.R.\ was supported by NSF CAREER Grant No.\ DMR-2238679 and is grateful to the Institute for Solid State Physics, University of Tokyo where a part of this work was performed for their hospitality. \\

\begin{center}
{\bf DATA AVAILABILITY}
\end{center}

No data were created or analyzed in this study.

\bibliography{ref}

\end{document}